\documentclass[12pt,preprint]{aastex}
\begin{document}

\newcommand{\kms}{km\ s$^{-1}$}
\newcommand{\cc}{cm$^{-3}$}
\newcommand{\co}{$^{12}$CO}
\newcommand{\thco}{$^{13}$CO}
\newcommand{\ceo}{C$^{18}$O}
\newcommand{\hco}{HCO$^+$}
\newcommand{\hthco}{H$^{13}$CO$^+$}
\newcommand{\cts}{C$^{34}$S}
\newcommand{\nh}{N$_2$H$^+$}
\newcommand{\vchan}{$v_{chan}$}
\newcommand{\vout}{$v_{out}$}
\newcommand{\venv}{$v_{env}$}

\title{The Evolution of Outflow-Envelope Interactions in Low-Mass Protostars}
\author{H\'ector G.~Arce\footnote{NSF Astronomy and Astrophysics
Postdoctoral Fellow}}
\affil{Department of Astrophysics,
American Museum of Natural History, New York, NY 10024} 
\email{harce@amnh.org} 
\and
\author{Anneila I.~Sargent}
\affil{Division of Physics, Mathematics, and Astronomy,
California Institute of Technology, MS 105-24, Pasadena, CA 91125}
\email{afs@astro.caltech.edu}

\shorttitle{Outflow-Envelope Interactions}
\shortauthors{Arce \& Sargent}

\begin{abstract}
We present multi-line and continuum observations
of the circumstellar environment within $10^4$ AU
of a sample of protostars to investigate how the effects
of outflows on their immediate environment 
changes over time. 
\co (1-0)  emission probes the 
high-velocity molecular outflows near the protostars
and demonstrate that the outflow opening angle widens
as the nascent star evolves. 
Maps of the
\thco(1-0)  and \hco(1-0) outflow emission 
show that
protostellar winds erode the circumstellar
envelope through the entrainment of the 
outer envelope gas. The spatial
and velocity distribution of the dense
circumstellar envelope, as well as its mass, is traced
by the   \ceo(1-0)  emission and also displays evolutionary
changes.  We show that outflows are largely
responsible for these changes, and propose an empirical
model for the evolution of outflow-envelope interactions.
In addition, some of the outflows in our sample
appear to affect the chemical composition of the surrounding
environment, enhancing the \hco \/ abundance.
Overall, our results confirm 
that outflows  play a major
role in the star formation process  through their strong physical
and chemical impacts on the environments of the young protostars.

\end{abstract}   


\section{Introduction}
\label{intro}

Young stellar objects
ranging in age from 
approximately $10^{3}$~yr to more than
$10^{6}$~yr are known to power outflows
that entrain  the surrounding
gas of the parent molecular cloud.
Protostellar winds originate close to the surface
of the forming star and must interact with the dense envelope
surrounding the protostar-disk system
(e.g., K\"onigl \& Pudritz 2000; 
Shu et al.~2000).  These circumstellar
envelopes, with sizes of
$10^{3}$ to 10$^{4}$ AU, are the primary
mass reservoirs for young stellar objects (YSOs).
As a result, 
perturbations caused when outflow 
momentum and energy are injected into the 
protostellar surroundings can have a major impact
on the mass-assembly process.
In fact, outflows are most likely to be 
responsible for the clearing of material from
the protostellar surroundings, a process that could
result in the termination of the infall phase 
(e.g., Velusamy \& Langer 1998) and 
affect the star-formation efficiency in the cloud
(e.g., Matzer \& McKee 2000).
However, there is still no general understanding
of how the outflow's effect  on its immediate
surroundings changes as the protostar evolves, as 
studies of the outflow-envelope interaction 
have been limited to a few specific cases.

In the nearest
star-forming regions,
circumstellar envelopes around low-mass young
stellar objects have  angular sizes
of about $10\arcsec$ to $30\arcsec$.
 A detailed 
study of the outflow impact on the envelope
requires observations that distinguish between
the various  velocity patterns 
that can exist in  
different density regimes in these envelopes
(i.e., infall, outflow, 
rotation, and turbulence).
Observations at high angular and velocity resolution 
of a variety of molecular species that trace a range 
of density and kinematic regimes are therefore essential.
Also, in order to examime how the outflow-envelope
interaction evolves, observations of the environment around 
protostars of different ages are needed.

Here, we present the results of a systematic
survey of the circumstellar environments 
of a sample of YSOs with outflows. 
Multi-line and continuum high spatial and spectral resolution 
observations from the Owens Valley  
Radio Observatory (OVRO) millimeter-wave  array enable
investigation of the outflows,  circumstellar gas and
evolution of the outflow-envelope interactions, on
scales of $\sim 10^3$ AU. Considerably more 
detailed studies of two of the sources in our sample, L\,1228 and RNO\,43,
have already been published 
(Arce \& Sargent 2004; 2005, hereafter Paper I and Paper II).

\section{The Sample}
\label{sample}

Our observed sample consists of  nine low-mass protostars 
at different evolutionary stages, and include: 1) three 
class 0 sources ---deeply embedded 
protostars in the earliest formation stage, with ages between 
$\sim 10^3$ and $\sim 10^4$~yr 
(see Andr\'e et al.~2000); 
2) three class I sources ---protostellar objects
still embedded in their parent cores but 
older than the class 0 sources, with ages $\sim 10^5$~yr
(see Lada 1999); and 3) three class II
sources, with
ages of a few $\sim 10^6$ yr (Lada 1999). 
All are within 500~pc of the Sun and are known 
outflow sources based on either  millimeter or optical
observations. Approximate protostellar ages are obtained from 
the bolometric temperature, $T_{bol}$, 
the temperature of a blackbody having the same mean frequency
as the observed spectral energy distribution (SED) of the YSO
(see Myers \& Ladd 1993; Chen et al. 1995; Ladd et al. 1998). 
Sources in our sample have $T_{bol}$ ranging from 40 to 2700~K,
corresponding to ages from  $\sim 10^3$ to $\sim 10^7$~yr.
 Table~\ref{sourcestab} lists the source names, 
coordinates of the observed fields, and other relevant parameters. 
A brief description for each source is given in the Appendix.

\section{Observations}
\label{obs}

The observations were carried out using the 
OVRO millimeter array of six 10.4~m telescopes, during 
three observing seasons between 2001 September and 2004 May. 
For all sources in our sample, a single pointing 
(i.e., no mosaicing) and 
two different correlation 
configurations were used to 
provide images on a total of six different molecular lines 
and two 4 GHz-wide continuum emission bands. The area on the sky
covered by each map is the OVRO primary beam,
 $65\arcsec$ $\times (100/\nu$) (FWHM), where $\nu$
is the observed frequency in GHz.
 
The first digital correlator configuration
provided simultaneous measurements of  
 $^{12}$CO(1-0), $^{13}$CO(1-0), and C$^{18}$O(1-0)   
at 115.27, 110.20, and 109.78~GHz, respectively, and
4~GHz bandwidth continuum emission at 2.7~mm.
With the local oscillator (LO) frequency set to 112.736~GHz, 
the $^{12}$CO line was observed in the upper sideband 
with a channel width of 0.325~\kms \/ over a  
20~\kms \/ bandwidth.
In the lower sideband, the \thco \/ and \ceo \/ lines 
were each observed
at a velocity resolution of 0.17~\kms \/ over
a 5~\kms \/ bandwidth.
Observations were obtained in two to five different 
array configurations depending on the source and with
baselines extending from 15 m to a maximum of 480 m. 

Simultaneous observations of 
HNC(1-0) F=2-1, HCO$^+$(1-0), and  H$^{13}$CO$^+$(1-0) 
at 90.66, 89.19, and 86.75~GHz, respectively, and
of  4~GHz bandwidth continuum emission centered 
at 3.4~mm were obtained with 
the LO frequency tuned to 87.971~GHz, so that the
HNC and \hco \/ lines were  in the upper sideband  and 
\hthco \/ in the lower.
All three lines were observed with two or three different 
array configurations with spectral
resolutions of $\sim 0.1$~\kms \/ over bandwidths of 6~\kms, 
and baselines from 15 m to a maximum of 120~m.
Only maps of the \hco \/ emission are discussed here, as 
they are relevant to the present discussion. 

A bright ($> 1$~Jy) quasar near each observed source was 
used  for phase and gain calibration. Flux calibration relied on 
observations of 3C273 and Uranus, with resulting uncertainties of 
about 20 to 30\%. Visibility data was edited and calibrated using the MMA
software package (Scoville et al.~1993) and images were produced
using the MIRIAD software package (Sault et al.~1995)
and its CLEAN algorithm, with ``robust'' weighting of the 
visibilities.  For each map the robust parameter was chosen
to optimize both spatial resolution and sensitivity
to extended structures, as well as the signal-to-noise ratio. 
Table~\ref{linestab} provides a
list of the relevant parameters for the molecular line
and continuum emission data.

\section{Results}
\label{res}
The gas surrounding forming stars is generally characterized by
complex kinematics and steep density gradients 
and observations of 
different molecular lines are necessary to trace different density and 
kinematic regimes. 
We use the optically thin 3 mm dust continuum emission 
to pinpoint the position of the sources in our sample and 
to study the inner circumstellar environment.
The $^{12}$CO(1-0) emission probes
the low-density,  high-velocity outflow gas, while the
less optically thick $^{13}$CO(1-0) reflects the behavior of
the low-velocity, slightly denser outflow, as well as the impact of the
protostellar outflow on the outer regions ($\gtrsim 2000$ AU)
of the circumstellar envelope. Observations of the C$^{18}$O(1-0) line 
trace the inner dense parts of the envelope surrounding the protostar
and in some cases
 may be used to probe the effect of the outflow on the high-density gas.
We originally observed HCO$^+$(1-0) with the expectation that 
this would trace
infall motions of the gas surrounding the young stars.  
 Since HCO$^+$(1-0) kinematics appear to be dominated
by outflow motions, we have used these results to examine
the physical and chemical impact of the outflow on the
surrounding gas. 

\subsection{Continuum Emission}
\label{cont}
Millimeter continuum thermal dust emission at 2.7 and 3.4 mm
was detected towards all the sources in our sample. 
Maps of this continuum emission were constructed 
 so that both bands had similar
synthesized beams  (see Table~\ref{linestab}).
Since the linear resolution of our continuum observations ranges from 
about 700 to 2700 AU, it is very probable that  both the emission
from the envelope around the protostar and that from an
unresolved circumstellar disk 
contribute to the total detected emission. We expect that in the
class 0 sources more than 80\% of the emission comes from the 
inner ($< 2000$~AU) circumstellar envelope.
For the class I and II sources, we expect
the contribution from the disk to the total
detected emission to rise from about 30\% to 90\% as the source
evolves 
(e.g., Hogerheijde et al.~1997a; Looney et al.~2000).

We estimate the total mass of the  
 gas and dust
 associated with 
the continuum emission, $M_{cont}$, from 
$F_{\nu} D^2 / B_{\nu}(T_d) \kappa_{\nu}$ (e.g.,
Beckwith et al.~1986).
Here, $F_{\nu}$ is the continuum flux density at frequency $\nu$,
$D$ the distance to the source, $B_{\nu}$ the blackbody (Planck) 
function, $T_d$ the dust temperature,
and $\kappa_{\nu}$ is the dust mass opacity coefficient, 
with the usual form, $\kappa_{\nu} = 0.1 (\nu/10^{12} ~ {\rm Hz})^{\beta}$ 
cm$^{2}$ g$^{-1}$ (e.g., Beckwith \& Sargent 1991). The
value of $\beta$  was determined by fitting the measured flux densities 
at wavelengths between 0.62 and 3.4 mm  with a power law,
$F_{\nu} \propto \nu^{(2+\beta)}$, using measurements from 
our observations and the literature 
(see references in Table~\ref{sourcestab}). For sources with less than 
three available flux measurements, we adopt $\beta = 1$. 
The dust temperature was estimated from greybody fits
to the spectral energy distribution of the source (e.g., Chini et al.~1997) 
at wavelengths greater than $60 \micron$ and using the 
estimated (or assumed) value of $\beta$. 
The measured fluxes at  2.7 and 3.4 mm, $F_{2.7}$ and $F_{3.4}$,
respectively, and the
estimated values of $\beta$, $T_d$, and $M_{cont}$ 
 are listed in Table~\ref{contab}.

\subsection{$^{12}$CO(1-0) outflows}
\label{twco_out}
In Figure~\ref{twcoouts} we present integrated intensity maps of the 
$^{12}$CO(1-0) emission associated with the observed sources.
The central envelope velocity,
$v_{env}$, and the velocity ranges of integration
are given in Tables~\ref{envtab} and \ref{velflowtab}, 
respectively.
Blueshifted and redshifted components are detected for all sources;
velocities deviate by at least 
 0.5~\kms  , and usually by more
than 1~\kms , from $v_{env}$. Since our 
interferometer observations filter out structures 
on scales greater than $\sim 30\arcsec$, we are confident that the
emission seen  in Figure~\ref{twcoouts} traces  molecular gas
accelerated by the opposing outflow lobes close to the 
protostellar sources, and not ambient material. 
However, filtering 
of large scale structures in our interferometer
 observations hampers our ability to {\it fully} recover the flux from the
 outflow gas present in the observed area. 

The amount of filtered or
 missing flux in our OVRO data depends on the observed molecular
 outflow's structure; a smaller fraction of the total flux
 will be recovered from molecular outflows composed of 
 smooth large-scale structures, compared to the fraction
 of the total flux recovered from molecular outflows 
 that  are mainly made of compact structures.
 For three sources
 in our sample, we compared
 published \co \/ data from the Five College Radio Astronomy (FCRAO) 14~m 
 telescope (Tafalla \& Myers 1997; Lee et al. 2000; Arce \& Goodman 2001)
 with our OVRO data averaged over a 
 $46\arcsec \times 46\arcsec$ area, the approximate beam size
 of the FCRAO \co \/ observations. We find that for L\,1228 our 
OVRO observations
recover more than
 60\% of the flux at all outflow velocities, and more than 90\% 
 of the flux at high outflow velocities. On the other hand, in HH\,300 and 
 RNO\,91 we recover 15 and 20\%, respectively,  of the single-dish flux
at outflow velocities.
 Evidently, the amount of missing flux is different for each molecular outflow. Here we do not
 attempt to correct for the missing flux in our OVRO data, as we lack the necessary single-dish
 data to do so for all sources.
 We stress that regardless of how much flux is missing in each of the 
 molecular outflow maps in our sample,  our data
  are sensitive to the small-scale outflow structures ---important for the study of the 
  outflow-envelope interactions--- 
  and our maps reliably show the 
 outflow morphology close to the source.

\subsection{$^{13}$CO(1-0) outflows}
\label{thcooutsec}
In Figure~\ref{thcoouts},  integrated intensity maps of 
the \thco \/  emission towards our sample
sources show gas that has been accelerated by 
the protostellar outflow and is slower and more dense than
the \co \/ features. 
  Like the \co  \/ features,
 the morphology of the 
\thco \/ blueshifted and redshifted outflow lobes varies substantially among sources.
Nevertheless, with the exception of the red \thco \/ lobe of GK/GI Tau, the 
locations and positional offsets of the \thco \/ and \co \/ outflow lobes 
correspond well. 
 In several sources the \thco \/ outflow emission is concentrated at 
 the  outer edges of the \co \/ outflows, and  appears to delineate the
 dense outflow cavity walls.
The blue lobes of L\,1228 and RNO\,129 are particularly nice examples of this
behavior but it is also suggested for RNO\,43. 
In other sources the \thco \/ outflow concentrates  
  at positions slightly offset from
 the protostar in the same direction as the corresponding
 \co \/ outflow lobes, and generally have opening angles
larger than that  seen in
 \co .  This morphology is best illustrated by HH\,114mms,
 HH\,300,  the red lobe of RNO\,129, and the blue lobes of T Tau 
and IRAS\,3282. Close to these 
sources, blueshifted and redshifted \thco \/ traces dense gas 
that has been propelled by the protostellar outflow, as well as entrained
 dense gas in the outer outflow walls. On ocassion,  \thco \/ is concentrated
outside the limits of the \co \/ outflow, as in the red lobe of T Tau and the
blue lobe of GK/GI Tau, perhaps due to the presence of a dense shell
produced by the interaction between the protostellar
outflow and the surrounding environment, and 
similar to the larger scale morphology
observed in other sources (e.g., Arce \& Goodman 2002a).

Similar to the \co \/ data, we expect that our \thco \/ interferometer maps 
filter a substantial amount of flux (see \S \ref{twco_out}). For example,
comparing single-dish (FCRAO) data of L\,1228 and HH\,300 
(Tafalla \& Myers 1997;  Arce \& Goodman 2001)
with our OVRO maps, we find that the interferometer \thco \/ observations
recover only 10\% and 15\%, respectively, of the flux in the outflow wings. This suggests that
our OVRO observations filter most of the \thco \/ (large-scale) cloud emission and 
the interferometer maps mainly show the (small-scale) 
emission associated with the envelope and the regions where it is being impacted
by the  underlying protostellar wind.



\subsection{Molecular Outflow Parameters}
\label{outflowpara}

The \co \/ line is expected to be optically thick 
at velocities close to the ambient envelope velocity 
 and optically thin at high outflow
velocities. Thus, 
we use \co \/ emission to estimate the mass of the low-density,
high-velocity component of the molecular outflow
and \thco \/
emission to account for higher density, lower velocity material
(see papers Papers I and II for more details).
At outflow velocities where the \co \/ line is
extremely optically thick and there is detectable 
resolved \thco \/ outflow emission, 
we derive the outflow mass from the
 \thco \/ emission, as it is
is expected to be optically thin
at these relatively high velocities 
far from the line core. Outside the velocity regimes where we use
\thco ,  we estimate the outflow mass from the \co 
\/ emission. 


We estimate the outflow mass, $M_{out}$, from
the \co \/ and \thco \/ data assuming 
LTE  and a gas temperature of 25~K. 
We adopt [H$_2$]/[\thco ] = $7 \times 10^5$
(Frerking et al.~1982) 
and [\thco ]/[\co ] = 62 (Langer \& Penzias 1993). 
We obtain the outflow momentum,
$P_{out} = \Sigma_{v} M_{out}(v) v$, and  kinetic energy, 
$E_{out} = \frac{1}{2} \Sigma_{v} M_{out}(v) v^2$,  for each source
and list their values in Table~\ref{outtabi}.
In addition, we calculate the mass outflow rate, $\dot{M}_{out} = \Sigma_{v} M_{out}(v) / \tau$, 
and the flow momentum rate, $F_{CO} = \Sigma_{v} M_{out}(v) v / \tau$ (see  Table~\ref{outtabii}),
where  $\tau = v/r$ is the dynamic time scale of  the outflow gas, $v$ is
the outflow velocity, and $r$ is the distance from the source to the outflow gas.
In Table~\ref{outtabii} we also show the dense gas mass outflow rate, 
$\dot{M}_{dense} = \Sigma_{v} M_{dense}(v) / \tau$, where $M_{dense}$ is the 
outflow mass traced solely by the \thco \/ emission.

For all mass-related outflow parameters listed in Tables~\ref{outtabi} and \ref{outtabii} 
(except for $\dot{M}_{dense}$, where we only use the \thco \/ emission)
the final estimate is the summation of the estimates 
from both the  \co \/
and  \thco \/ emission. 
These values should be treated as lower limits, since no correction for missing flux, 
opacity or for outflow inclination angle have been applied.
Care should also be taken when comparing $M_{out}$, $P_{out}$, and $E_{out}$
for the individual outflows
 since the areas observed are in effect,  the
linear sizes of the interferometer primary beam, and differ according to source 
distance. Nevertheless, to the first order $\dot{M}_{out}$, 
$F_{CO}$, and  $\dot{M}_{dense}$ 
do not depend on the observed area (Henning \& Launhardt 1998).

\subsection{HCO$^+$ outflows}
\label{hcooutsec}
In Figure~\ref{hcoout} the  \hco \/ emission at 
$\mid v_{env} \mid > 0.5$~\kms \/ appears to
coincide with 
the molecular outflow lobes. Indeed, outflowing
\hco(1-0) is observed in all sources except  
GK/GI Tau, although not all show both blueshifted and 
redshifted emission.
In general,
this \hco \/ is confined to the outflow cavity walls.
In this respect it echoes the behavior of \thco \/ but,
as can be seen in Table~\ref{velflowtab}, 
the velocity range exceeds that of the \thco \/ outflow
emission in most sources, and, as can be seen in 
Figure~\ref{hcoout},
in some sources the \hco \/ outflow emission does not trace
the full extent of the outflow cavity walls. 
As in  the \thco \/ data, we expect that 
our interferometer observations have filtered much of the
extended emission, and that the \hco \/ outflow maps show 
the compact structures in the envelope where the gas has 
been influenced by the prostellar outflow.

The \hco \/ abundance in the outflow,  
$X_{HCO^+}$, 
should indicate whether outflow-triggered chemical
processes have affected the surrounding gas.
The \thco \/ outflow emission  along the outflow cavity walls
and offset from the source position is probably optically thin,
since there is no \ceo \/ at these velocities and positions. 
Thus, it provides a reliable estimate of 
the molecular hydrogen column density, $N_{H_2}$, using the same 
assumptions as in \S~\ref{outflowpara}. 
The \hco \/ column density, $N_{HCO^+}$, can be obtained assuming the
emission is optically thin, and leads to the outflow
\hco \/ abundance using $X_{HCO^+} = N_{HCO^+}/N_{H_2}$, where
$N_{H_2}$ (from the \thco \/ data) and $N_{HCO^+}$ are obtained over
the exact same area and velocity range. 
The uv-coverage and beam sizes of our  \thco \/
and \hco \/  data are very similar. Thus, these maps are sensitive to similar 
structures in the emission and may be used to obtain 
a reliable estimate of the  \hco \/ abundance in the outflow
and outflow-envelope interface, close to the protostar.  
Given that an estimate of
$X_{HCO^+}$ can only be obtained
in regions where the \hco \/ and \thco \/ emission coincide
in position and velocity, and that in some sources the \hco \/
emission aries from positions and velocities where \thco \/
is not detected, we can obtain only a lower limit to
$X_{HCO^+}$ for most of our sources (see Table~\ref{hcotab}).

\subsection{C$^{18}$O(1-0) envelopes}
\label{ceoenv}
Integrated intensity maps of the \ceo \/ emission associated with 
our sample are shown in Figure~\ref{ceocores}.  For
the class 0 and class I sources, \ceo \/ emission
usually coincides with continuum peak positions. 
Elliptical gaussian fits indicate geometric mean sizes  
(after beam deconvolution)
of about 2000 AU (FWHM) independent of evolutionary stage.
An exception is RNO\,43 for which the \ceo \/ emission region 
size is about 5600 AU (FWHM) 
 (see Table~\ref{envtab}).  
The compact morphology and spatial
coincidence with the dust continuum peaks strongly
 suggest the  \ceo \/ emission 
traces the  dense circumstellar
 envelope gas in these class 0 and
 I sources. By contrast, \ceo \/ emission 
is significantly offset from the
positions of our class II sources, and may
 trace the remnants of the 
high-density ($n \gtrsim 10^{4}$~cm$^{-3}$) circumstellar
 envelope gas that has been mostly cleared by the 
protostellar outflow (see \S \ref{outenv}).
  
Comparing the total \ceo \/ 
flux obtained from single-dish (FCRAO) data of two sources 
(HH\,300: Arce \& Goodman 2001; IRAS\,3282: Hatchell et al.~2005), with the 
flux recovered by our interferometer data (smoothed to the same
beam size), we estimate that the OVRO  observations
trace 1 to 4\% of the total flux traced by the 
FCRAO observations. In addition, the velocity distribution of the emission in the
single-dish data is very different from that of our OVRO data. 
These properties support the idea that the two sets of data trace
entirely different structures: 
the \ceo \/ emission detected in our
OVRO data arises from the compact structure of dense gas surrounding the
protostar (i.e., the circumstellar envelope, or its remnants), 
while the single-dish \ceo \/ observations 
trace the surrounding (large-scale) cloud. Thus, our \ceo \/
OVRO data can be
used to obtain an estimate of the circumstellar envelope mass
(e.g., Hogerheijde et al.~1998).

We estimate the masses of the dense circumstellar envelope, 
$M_{env}$, assuming optical thin \ceo \/ emission, LTE, and an
abundance ratio of [\ceo ]/[H$_2$] = $1.7 \times 10^{-7}$ 
(Frerking et al.~1982). 
We adopt an excitation temperature, $T_{ex}$, of 25~K, since
$T_{ex}$  probably lies between
10 to 40~K; the lower limit is given by the
``typical'' temperature of an isolated dense core 
(Jijina et al.~1999), while 
the upper is the ``typical''  $T_d$  of our sources.
Envelope masses and other properties are listed
in Table~\ref{envtab}.

\section{Discussion}
\label{discuss}

\subsection{Outflow Opening Angle}
\label{outflowsec}

The CO outflows displayed in 
Figure~\ref{twcoouts} have very different morphologies, due in part
to their various surrounding environments. Nevertheless, there are
significant trends. The molecular outflows from the
youngest protostellar sources (class 0) tend to be highly collimated 
with either jet-like morphologies or cone-shaped
lobes with relatively small opening angles of less than 55\arcdeg . 
As discussed in Paper II, the morphology and kinematics of one of these,
the RNO\,43 outflow, strongly suggest that it results from  the 
entrainment of ambient gas by an underlying collimated jet.
Likewise, the significant collimation of the 
HH\,114mms molecular outflow, as well as the fact that
the outflow velocity increases with distance from the
source, indicate that it is also driven by a jet or by a
very collimated, momentum-conserving, wind 
(e.g., Lee et al.~2001).  The cone-shaped morphology
of the IRAS\,3282 outflow resembles that 
observed in other class 0 outflows such as L\,1448
 (Bachiller et al.~1995),  L\,1157 (Gueth et al.~1996), and
HH\,211 (Gueth \& Guilloteau 1999). These cone-shaped morphologies have
been successfully fitted with  jet bow shock entrainment models
(e.g., Raga \& Cabrit 1993). 
By analogy, we propose that the
 outflow lobes in IRAS\,3282 represent the walls
of a cavity produced by a jet bow shock. The
limited velocity range of our \co \/ observations (see Table~\ref{velflowtab})
precludes our detecting the high velocity collimated component of IRAS\,3282
that has been observed in single dish maps  
at outflow velocities of $\sim 50$~\kms \/ by Bachiller et al.~(1991).

The sightly more evolved (class I) sources  in our sample have
\co \/ outflows  with 
lobe opening angles of more than 75\arcdeg . 
This wide angle morphology is also seen in \thco \/
(Figure~\ref{thcoouts}) with opening angles for 
L\,1228, HH\,300, and RNO\,129, which are as 
wide as, if not wider than, their \co \/ counterparts.
Near-infrared and optical scattered light nebulae 
associated with these sources (Bally et al.~1995; Reipurth et al.~2000;  
Movsessian \& Magakian 2004) display morphologies that are very
similar to those of the molecular outflow lobes, consistent with the high-velocity
\thco \/ emission tracing the walls of cavities produced by the protostellar
outflow.

By contrast, outflows from the most evolved young stars in 
our sample, the class II sources, have
no definite shape or structure. Blueshifted and redshifted
gas within about 6000~AU of 
RNO\,91, T Tau, and GK/GI Tau shows neither
the collimated appearance
of the class 0 outflows or the wide-angle morphology characteristic
of the class I sources.
Class II sources are generally found in regions
of low ambient density (Myers et al. 1987), and the clumpy structures seen
in Figure~\ref{twcoouts} may arise only
in the (limited) regions where 
molecular gas around the nascent star has not yet
been cleared by the protostellar outflow.
RNO\,91 is a case in point. A large scale 
mosaic 
of \co \/ interferometer observations,
covering an area 
about $5 \times 10^4$ by $10^5$ AU,
demonstrates that the outflow
has produced a $3 \times 10^4$~AU wide cavity, with an opening
angle of 160\arcdeg \/ (Lee et al.~2002; Lee \& Ho 2005). 
Thus, our observations of the RNO\,91 source, with a primary beam
of $\sim 5 \times 10^3$~AU,
cannot fully map the cavity walls seen in the maps of Lee et al.~(2002), 
instead
we can expect to detect only remnant molecular material 
that has not been swept up out of the field of view of our maps. 
A similar scenario 
applies for T Tau (Momose et al.~1996)
and perhaps for GK/GI Tau.

In Figure~\ref{openang} we plot 
the opening angle, $\theta$, of each of our 
observed  outflows as a function of the
source age, together with published data of other  
spatially resolved molecular outflows from low-mass YSOs, within
500~pc of the Sun, where the outflow opening angle near the source
could easily be measured (see Table~\ref{openangtab}).
In cases where the value of $\theta$ 
depends on velocity, we have adopted the 
largest opening angle.
It is immediately clear that there is a tendency
for the outflow opening angle to increase with the
evolutionary stage of the outflow source. A fit to our data 
alone yields $\rm{log}(\theta/\rm{deg}) = (1.1 \pm 0.2) + 
(0.16 \pm 0.4) \rm{log}(t/\rm{yr})$, quite consistent
with the fit
to the combined data,  $\rm{log}(\theta/\rm{deg}) = (0.7 \pm 0.2) + 
(0.26 \pm 0.4) \rm{log}(t/\rm{yr})$.

Although an increase of outflow opening angle with age 
has been inferred by other authors
(e.g., Richer et al.~2000), it has never been clear if the trend was 
a real effect or due to the varying beam sizes in independent 
studies of individual sources observed at lower resolution
single telescopes. Indeed, 
high-resolution observations indicated that the
opening angle of the B5-IRS1 outflow is widening at about 
0.006\arcdeg \/ yr$^{-1}$ (Velusamy \& Langer 1998), but there
was no suggesting evidence to confirm 
that this was a general trend. It is true that interferometer
mapping of  a sample of molecular 
outflows from sources at different 
evolutionary stages 
concluded that outflow {\it widths} increase with age (Lee et al.~2002).
 However, we argue that the outflow opening angle is a better probe of 
the outflow's impact on the envelope, since it is measured
close to the protostar, where
the outflow-envelope interactions are strongest.
Outflow width, by contrast, may be measured anywhere along
the outflow length, including regions far
away from the circumstellar envelope.
Our results demonstrate unambiguously that outflow opening angles
increase with source age, implying a progressively stronger outflow-envelope
interaction.

\subsubsection{Why should we expect outflow widening?}

There is as yet no generally accepted  explanation for the observed
outflow widening.  It has been proposed that
jet axis wandering (precession) could produce wider cavities as the
protostar evolves (Masson \& Chernin 1993).
However, although ample evidence for 
jet wandering exists (e.g., Reipurth et al.~1997; Terquem et al.~1999), in most sources
the opening angle of the precession cone is smaller that the
observed wide-angle outflow cavity (e.g.,  Yu et al.~1999;
Reipurth et al. 2000; Paper I). Our study of the L\,1228 outflow
also shows that the detailed
morphology of the
wide-angle cavities is incompatible with their
being produced by jet entrainment (Paper I).  
An attractive  alternative is an underlying stellar wind
with both a collimated and a wide-angle component
(e.g., Kwan \& Tademaru 1995;
K\"onigl \& Pudritz 2000, Shu et al.~2000; Matt et al.~2003)
where 
 the observed molecular outflow
is predominantly  driven by one of the two components
depending on the age of the protostar.
The X-wind model of Shang et al.~(1998), for example, 
predicts a stellar wind with a high
density along the axis (very similar to a jet) but decreasing 
density at increasing angles from the axis. 
It is conceivable that at very early ages only the dense collimated
part of the wind can break out of the  
surrounding dense (infalling) envelope 
(e.g., Wilkin \& Stahler 2003), thereby producing the jet-like 
outflows observed in class 0 sources.
As the envelope 
loses mass, 
through infall onto the star-disk system {\it and} outflow
entrainment along the axis (see \S~\ref{outenv}), the less
dense and wider wind component will  break through, 
entraining the gas unaffected by the collimated component.
At this stage there will be almost no gas along the outflow 
axis, and the gas entrained by the wide-angle component of the wind
will be the dominant structure in the molecular outflow,
as seen in class I sources. 
 A decrease in infall as the protostar evolves 
(e.g., Henriksen et al.~1997) 
could also allow the wind to carve out wide-angle cavities 
(see Delamarter et al.~2000). 



\subsection{Evolution in the Envelope's Morphology and Kinematics}
\label{envmorph}

There also appears to be a change in the \ceo \/ morphology as the
outflow source evolves. From Tables~\ref{outtabi} and \ref{envtab}
it is clear that for class 0 sources the \ceo \/ is usually elongated
along an axis within 15\arcdeg \/ of that of the outflow.
In Figure~\ref{envgrad}, two Class 0 sources, RNO\,43 and IRAS\,3282,
display velocity gradients along the (elongated) major axis of the envelopes,
in the same sense as the outflow. There is no clear velocity gradient
in the \ceo \/ envelope of HH\,114mms, although the gas at
the extreme southwest edge of the envelope  is blueshifted 
---similar to the  CO outflow. Based on our detailed analysis 
of \ceo \/ emission in RNO\,43 (Paper II), we suggest that the outflow is 
responsible for the velocity field and morphology of the high-density
envelope around at least two class 0 sources in our sample. 

On the other hand, Figure~\ref{envgrad} and 
Tables~\ref{outtabi} and \ref{envtab} indicate
the \ceo \/ envelopes around the class I sources 
tend to be elongated perpendicular to the outflow axis,
and there is no obviously shared velocity distribution. 
For both RNO\,129 and L\,1228 the \ceo \/ velocity gradient is 
aligned with the major axis of the envelope and 
perpendicular to the outflow axis, consistent with 
rotation. In HH\,300, the envelope 
 velocity 
gradient is along the {\it minor} axis, 
parallel to the outflow axis, suggesting
a flattened infalling 
envelope, where infall motions produce blueshifted
and redshifted emission in the far and near sides of
the structure, respectively (as seen in other
sources, e.g.,  Torrelles et al.~1995;
Ohashi et al.~1997; Momose et al.~1998). Contrary to
what is observe in the class
0 sources shown here and in the literature where the 
outflow is clearly impacting the dense circumstellar 
gas (i.e., Gueth et al.~1997; Wiseman et al.~2001;
Beltr\'an et al.~2004), the 
\ceo \/ emission in HH\,300 is elongated 
perpendicular to the outflow at all velocities, and does not
show any protrusion along the outflow axis.
We are, therefore, certain that the HH\,300 velocity
distribution is mainly (if not all) due to infall.

For class II sources,
either no \ceo \/ emission is detected or the emission
is offset in both position and velocity from the protostar.
Specifically, the \ceo \/ emission associated with RNO\,91 is
 about 8\arcsec \/ (1300 AU) southwest of the
continuum peak and is seen only at  velocities that are
blueshifted
with respect to that of the ambient envelope velocity. 
Similarly, the \ceo \/ emission associated with T Tau is
concentrated 15\arcsec \/ (2100 AU) east of the young star and
most of the emission emerges at velocities 
blueshifted by 0.2 to 0.5~\kms
\/ with respect to $v_{env}$. In both cases  
the \ceo \/ emission is concentrated toward the edge of the 
outflowing  \co \/ structure seen in Figure~\ref{twcoouts}. 
No \ceo \/ emission within our sensitivity levels was
detected from GK/GI Tau.

As we describe in \S\ref{outenv},
we interpret these changes in the \ceo \/ 
envelope properties between class 0 and class II sources
as a real evolutionary
progression due to the protostellar outflow's effect on
 the dense circumstellar gas.

\subsection{Envelope Mass Evolution}
\label{envmas}


In Figure~\ref{envmasfig} we plot our estimate of the 
envelope mass derived from the \ceo \/ emission, $M_{env}$,
as a function of source age, determined from 
the $T_{bol}$-age relationship of Ladd et al.~(1998).
Although the correlation is not very tight, $M_{env}$ is 
certainly greater for younger protostars
than for more evolved YSOs. We would not expect
any relation between $T_{bol}$ and $M_{env}$ a priori 
since $T_{bol}$ is obtained 
from the SED, and thus is an indirect
measure of the dust content close to the protostar 
rather than the gas. 

A linear fit to the log($M_{env}$)-log(age) plot results in:
\begin{equation}
{\rm{log}}(M_{env}/{\rm M}_{\sun}) =
 (0.87 \pm 0.5) - (0.45 \pm 0.09){\rm{log}}(t/\rm{yr}),
\end{equation}
with a correlation coefficient of 0.8. Taking the derivative, leads
to a rough estimate of the envelope mass-loss rate, $\dot{M}_{env}$, from:
\begin{equation}
\dot{M}_{env} \sim - 3t^{-1.45} 
\end{equation}
Evidently, $\dot{M}_{env}$ decreases rapidly with time from 
$\dot{M}_{env} \sim 10^{-4}$~M$_{\sun}$~yr$^{-1}$ for sources with
age $10^3$ yrs, to  
$5 \times 10^{-6}$, and 
$2 \times 10^{-7}$~M$_{\sun}$~yr$^{-1}$ 
for ages of $10^4$, and $10^5$~yrs,
respectively.  This is consistent with the core mass-loss
rate found by Ladd et al.~(1998), $\dot{M} \propto t^{-1.52\pm 0.55}$,
from lower resolution  \ceo (1-0)  observations of a sample of
Taurus cores harboring deeply embedded YSOs. 

Does the protostellar outflow play a role in the mass-loss of the circumstellar
envelope? As discussed in \S\ref{thcooutsec} (and in Papers I and II in more detail),
high-velocity  \thco \/ traces
dense envelope gas that has been entrained by the protostellar wind 
and thus provide an estimate of the outflow rate of the relatively
dense gas that  could escape the envelope's
gravitational potential. The dense gas mass outflow rate,
$\dot{M}_{dense}$, shown in Table~\ref{outtabii} is 
approximately constant across our sample, mostly ranging from
only 1 to about $5 \times 10^{-6}$~M$_{\sun}$~yr$^{-1}$, independent of
the source's age.
Only two sources, IRAS\,3282 and GK/GI Tau, deviate, 
with  $\dot{M}_{dense}$ values of $\sim 4 $
and $2 \times 10^{-7}$~M$_{\sun}$~yr$^{-1}$, respectively.
The median (and average) value of $\dot{M}_{dense}$ from our whole sample is 
between 1 and $2
 \times 10^{-6}$~M$_{\sun}$~yr$^{-1}$.
This is very close to the value of the envelope
mass-loss rate 
for late class 0/early class I sources with ages of about $10^4$ yrs, 
 $ \dot{M}_{env} \sim 5 \times 10^{-6}$~M$_{\sun}$~yr$^{-1}$ (see Equation 2).
It is, therefore, conceivable that a fraction of the mass-loss in the envelope is due to the
outflow entrainment of dense envelope gas.
In class I (and older) sources, with ages $> 10^5$ yr, 
the envelope mass-loss rate is even less, 
$ \dot{M}_{env} < 2 \times 10^{-7}$~M$_{\sun}$~yr$^{-1}$, and is about a factor
of ten smaller than the average $\dot{M}_{dense}$ in our sample. 
Therefore, in class I (and older) sources
even if only 10 to 20\% of the dense mass 
entrained by the outflow eventually escapes from the
envelope's gravitational potential, the outflow-envelope interaction 
can contribute significantly to the envelope's mass loss.

We note that in sources  
younger than $\sim 10^{3}$~yr the 
outflow may be responsible for only a small fraction of 
the total mass-loss in the envelope since
the estimated dense gas outflow mass rate is less than 10\% of 
the estimated envelope mass-loss rate.
This accords well with theoretical and observational work 
suggesting that  
during this very early stage of protostellar evolution
the envelope mass loss rate is dominated
by gas infall onto the star-disk system (e.g., 
Ladd et al.~1998; Henriksen et al.~1997;
Bontemps et al. 1996).

\subsection{Outflow-Envelope Interactions: An Empirical Model}
\label{outenv}

The widening of outflow cavities, the change in 
circumstellar envelope morphology and kinematics,
and the decrease of circumstellar mass 
and mass-loss rate with source
age all indicate that the protostellar outflow 
has a major role in the early stages of star formation,
particularly in the
evolution of the dense circumstellar envelope.
Our data support the following  empirical
model of the evolution of outflow-envelope interactions
(shown schematically in Figure~\ref{modelfig}). 

At very early stages of protostellar 
evolution, the young, powerful, and well collimated outflow
entrains and pushes on gas as it
burrows through the envelope, altering  
the distribution and kinematics of dense gas. 
This interaction causes the elongated 
distribution of the dense gas, as traced by the 
\ceo  \/ in our observations, 
 and a velocity gradient along the outflow axis
on scales of $10^3$~AU (see Figure~\ref{modelfig}).  
At this early stage the envelope mass-loss rate is
dominated by the infalling envelope gas 
onto the star-disk system.

With time, the envelope mass decreases and
the outflow cavities widen. This enables  
circumstellar gas at larger angles from the 
outflow axis to be entrained and cleared away
by the outflow until,
eventually, most of 
the dense circumstellar gas is concentrated
outside the outflow cavities in structures 
 perpendicular to the outflow axis, thereby
decreasing the  volume of the infalling region
(see Figure~\ref{modelfig}).
At this stage
most of the remaining dense circumstellar gas
has not yet been perturbed by outflow
motions, and the velocity structure 
of the envelope is dominated by 
either infall or rotation,
depending on the age of the source 
(see Hogerheijde 2001). 
The  
outflow-envelope interaction is mainly traced
by the \thco \/ outflow emission, as the 
lower density in the outer envelope precludes
the use of \ceo \/ as a probe. From this point
(class I and later), the mass-loss rate of the envelope
is dominated by the outflow's clearing effect.

As the nascent star evolves further,
the outflow lobes keep widening and by the
time the source reaches class II status most
of the dense circumstellar envelope has been
swept away by the outflow or has
fallen into the circumstellar disk-star system.
Only remnants of the original protostellar 
environment will be detectable 
(see Figure~\ref{modelfig}). 
In some cases, gas that has been
entrained by the outflow may form relatively dense 
shells and clumps at the edge of the outflow lobe,
that can be observed in \thco \/ and \ceo \/ emission
around class II sources  (as seen in our sample
and in  Welch et al. 2000).
We speculate that as the YSO
evolves further and reaches the end of the
class II stage the very wide outflow cavities will finish
clearing the surrounding gas that is left,  
halting the infall process altogether.

\subsection{Outflow Chemistry}
\label{outchem}

Our multi-molecular line observations also provide evidence that 
protostellar outflows can have a chemical impact on their envelope.
As discussed in \S~\ref{hcooutsec} and shown in Table~\ref{hcotab},
we can estimate the \hco \/ abundance, X$_{HCO^+}$, in the outflow. 
Comparing our results with a ``standard'' abundance of 
$\sim 7 \times 10^{-9}$ (J{\o}rgensen et al.~2004) we see that  
the \hco \/ abundance is  clearly enhanced in the 
IRAS\,3282 and HH\,114mms  outflows. 
In the case of L\,1228, the lower limit of
X$_{HCO^+}$ in the blue and red outflow lobes 
is similar to, and less than a factor of two less 
than, the ``standard'' \hco \/ abundance, respectively.
In the case of RNO\,43, 
the estimated  X$_{HCO^+}$ lower limit of the RNO\,43 outflow
is similar to the ``average'' 
 \hco \/ abundance in envelopes around Class 0 sources 
found by J{\o}rgensen et al.~(2004). It is possible
that in these two sources the \hco \/ abundance
may be slightly enhanced, but our data is not 
sufficient to firmly conclude this.
In the other sources in our sample,
the \hco \/ is similar to, or below, the standard abundance.

The \hco \/ emission 
seen in Figure~\ref{hcoout}  arises mainly along the outflow 
cavity walls or  the outflow lobes, and the observed
overabundance is consistent 
with  chemical models that  predict  
\hco \/ enhancement in  outflow
walls due to the shock-triggered 
release of molecules from dust mantles
and  subsequent chemical reactions in the warm gas
(Rawlings et al.~2000;
Viti et al.~2002;
Rawlings et al.~2004).  Hence, it is clear that
some of the sources in
our sample power outflows that are affecting 
the chemical composition of their surroundings.
On the other hand, \hco \/ outflows with abundances at or below
the standard value must arise in
 dense ($n \gtrsim 10^3$) envelope gas (see Evans 1999)  
that has been entrained by the outflow, where shocks have not
seriously affected the chemical properties of the environment.
In some sources the \hco \/ outflow emission does not trace 
the entire extent of the outflow wall (e.g., HH\,300, L\,1228)
or does not trace both outflow lobes (e.g., HH\,114mms, RNO\,43).
This could result from  differences in the environmental
conditions such as density, shock-induced radiation, 
or chemical composition, or to the effects of optical depth.

 \section{Summary}
Our systematic,
high angular resolution (3\arcsec \/ to 7\arcsec ),
multi-molecular line, survey of the circumstellar 
environment within $\sim 10^4$ AU of nine protostars at 
different evolutionary stages has enabled  
 detailed studies of the kinematics, morphology, density distribution,
and chemistry of the circumstellar gas. These, in turn have permitted
investigations of how the circumstellar envelopes and protostellar
outflows change with time.

Our
 \co \/ images trace high-velocity outflowing  molecular
material around all sources.
We detect a clear trend in the morphology of the 
protostellar molecular outflows. The youngest
sources (class 0) in our sample power
 molecular outflows with jet-like morphologies or
cone-shaped lobes with opening angles less than 
55\arcdeg .
The sightly more evolved (class I) sources  
in our sample have molecular outflows  with 
lobe opening angles of more than 75\arcdeg .
Outflows from the most evolved young stars in 
our sample, the class II sources, have
even wider lobes 
or no definite shape or structure.
Combining our 
data  with that from a number of sources in 
the literature shows that 
outflow opening angle close to the source
widens with time.

The denser, lower velocity gas probed by our \thco \/ and \hco
observations appears to arise from  the outflow cavity walls or very
close to the protostar. We suggest that this is dense 
gas from the outer regions of the circumstellar envelope that has been
entrained by the high velocity flow, thereby eroding the envelope and 
helping to widen the outflow cavities. We also suggest that 
evolutionary changes in the morphology and velocity field of the 
dense circumstelar envelopes, detected in our
 \ceo \/ images, are mainly caused by 
outflow-envelope interactions.
Class 0 source envelopes, for example, are  elongated with 
velocity gradients 
along the outflow axis, indicating that the 
young and powerful protostellar outflows  can entrain
dense envelope gas. Class I envelopes by contrast are elongated
more or less perpendicular to the outflow axis and concentrated
outside the outflow lobes, as if 
 most of the dense gas along the outflow axis 
has already been displaced.
 By the class II stage the sparse   
\ceo \/ emission observed is constrained to the outflow lobe edges,
consistent with most, if not all, of the dense gas being cleared.
 
Not unexpectedly, our results show a decrease in envelope mass 
as the age of the protostar increases. Comparisons of the estimated 
envelope mass-loss rate with the dense gas outflow rate imply
that the protostellar outflow plays an important role in 
envelope mass-loss during and after the class I stage. 
For younger protostars,
envelope mass-loss 
is more likely to be dominated by large infall rates.
Finally, enhanced abundance  of  \hco \/ in the outflow
in some of the sources in our sample indicate that
 the chemical
composition of the environment around these 
protostars is affected by shock-induced chemical processes 
 in the outflow-envelope interface.

Taken together, these trends demonstrate that outflows 
from protostars have a major physical
and chemical impact on their environments throughout
the entire star formation process and are an integral part
of that process. An understanding of their origins, evolution, 
and detailed characteristics is clearly critical for theories
of star formation.

 \acknowledgments
H.G.A. is supported by an NSF
Astronomy and Astrophysics Postdoctoral Fellowship under
award AST-0401568.
The Owens Valley Radio Observatory millimeter array is supported by the NSF 
grant AST-0228955. 
We are greateful to Chin-Fei Lee, Mario Tafalla 
and Jennifer Hatchell for sharing their single-dish data with us. 

\appendix
\section{Description of Sources}

\subsection{HH\,114mms}
The HH\,114mms protostar is a Class 0 source in L1589, a cloud
in the western part of the $\lambda$ Ori molecular shell. It 
was first detected at 1.3~mm 
(Chini et al.~1997), and has also been observed at 3.6~cm 
(Rodr\'{i}guez \& Reipurth 1996).
The \co(1-0) map of the nearby HH\,114/115 molecular outflow by  
Lee et al.~(2002)
partially overlaps the blue lobe of the HH\,114mms CO outflow.
Our data show the HH\,114mms CO outflow
is highly collimated, with a small bend in both outflow lobes 
about  7\arcsec \/ away from the source, where the outflow 
axis appears to change direction. This morphology may indicate
that the axis of the underlying protostellar wind has changed
over time. 
CO channel maps show that in both outflow
lobes the highest velocity gas is concentrated farthest from the source. In addition,
the opening angle of the CO blue lobe  widens slightly at lower outflow velocities, 
rather like the HH\,211 outflow (Gueth \& Guilloteau 1999).
The morphology and velocity distribution of the outflowing gas suggests
that an underlying jet-like wind
 drives the molecular outflow. 

\subsection{RNO\,43}
A detailed study of the RNO\,43 molecular outflow,
the circumstellar envelope and the
outflow-envelope interactions 
within $2 \times 10^4$ AU
of the protostar is presented in Paper II.
This outflow lies
in the L 1582B dark cloud at a distance
of about 460~pc,
in the $\lambda$ Ori ring
(Bence et al.~1996). 
The outflow arises from  IRAS\,05295+1247, a deeply embedded
class 0 source that
has been observed from far-infrared to centimeter wavelengths
(Cohen et al.~1985; Beichman et al.~1986;
Zinnecker et al.~1992; Reipurth et al.~1993; 
Anglada et al.~1992; Chini et al.~1997).
The 3.4~pc long RNO\,43 optical outflow 
includes four major Herbig-Haro (HH) knots,
HH 243, HH 244, HH 245, and HH 179
(Cohen 1980; Reipurth et al.~1997). The large-scale   
molecular (CO) outflow
extends about half a degree ($\sim 4$~pc) on the sky 
(Cabrit et al.~1988; Bence et al.~1996).
The positions of the HH knots and the morphology of the
giant molecular outflow indicate that the axis of the RNO\,43 flow
has changed over time (Bence et al.~1996; Reipurth et al.~1997).

\subsection{IRAS\,3282}
Single dish \co(2-1) observations of the environment surrounding 
IRAS\,3282 by Bachiller et al.~(1991) revealed a young, 0.5~pc long,
fast, and highly-collimated bipolar molecular outflow. 
The high velocity outflow ($\sim 50$~\kms ) gas
has a jet-like morphology, while the slower ($\sim 10$~\kms ) component
 exhibits wider outflow lobes.
Observations of the ammonia core suggest that the dense core
gas is being accelerated along the outflow cavity walls (Tafalla et al.~1993). 
High resolution OVRO observations of the 1.3~mm dust continuum,
reveal that IRAS\,3282 is a binary source with a 420~AU
separation and a mass ratio of 0.2 (Launhardt 2004).
Our \co(1-0) data  show that at outflow velocities between about 
2 and 7~\kms \/ both lobes have a V-shaped morphology, consistent
with a jet bow shock entrainment model (e.g., Gueth et al.~1996;
Gueth \& Guilloteau 1999).

\subsection{HH\,300}
IRAS\,04239+2436, the source of the HH\,300 outflow, is located in the westernmost
region of the B18 cloud  (sometimes refereed as the B18w region) in the Taurus molecular
cloud complex. Infrared photometry and spectra indicate that the source is a deeply
embedded class I protostar (Myers et al.~1987; Greene \& Lada 1996). 
Dust continuum emission surrounding the
source was mapped at 1.3~mm (at $\sim 12$\arcsec \/ resolution)
by Motte \& Andr\'e (2001).
The HH\,300 optical outflow is about 1.2~pc long and comprises three different
HH knots (Reipurth et al.~1997). 
The density, velocity and momentum distribution of the large-scale molecular
outflow indicates that it is driven by an underlying precessing and episodic wind
(Arce \& Goodman 2001).
HST images at 1.644 and 2.122 \micron \/ by Reipurth et al.~(2000)
indicate that IRAS\,04239+2436 is a binary source. In addition, they reveal 
the existence of a parabolic nebula with a small (1000 AU) jet along its symmetry 
axis, northeast of the source (i.e., the blueshifted side of the outflow). 
In our \co \/ images, the blue and red molecular outflow lobes have  shapes
and orientations similar to that of the near-infrared parabolic nebula.

\subsection{L\,1228}
A detailed 
 study of the infalling envelope, the wide-angle 
L\,1228 molecular outflow, and its 
impact on the circumstellar 
envelope is presented in Paper I.
This outflow is named after its parent
molecular cloud, located in the 
Cepheus flare  cloud complex at a 
 distance
of 200~pc (Kun 1998). The outflow source,
IRAS\,20582+7724, is an
an embedded (Class I) low-luminosity
protostar associated with a 
$\sim 1$~pc-long bipolar
molecular outflow, first observed in CO(1-0) 
by Haikala \& Laureijs (1989). IRAS\,20582+7724 also powers a 
$\sim 1.5$~pc-long optical flow, HH\,199 
(Bally et al.~1995).
The single-dish, multi-molecular line study
by Tafalla \& Myers (1997) indicates that the dense
core gas is being disrupted by the L\,1228 outflow. 
The tentative detection of a binary companion is reported in Paper I.

\subsection{RNO\,129}
RNO\,129 is located about 1\arcdeg \/ north of the L\,1228 core and is
believed to be associated with the same cloud (Kun 1998). 
The protostellar
source associated with the RNO\,129 nebulosity is IRAS\,21004+7811
(Ogura \& Sato 1990).  An extensive recent
 study of the region  by Movsessian \& Magakian (2004)
 presents optical images of the RNO\,129 nebulosity, and the HH objects
 near it, and concludes that  the central star 
 is a binary that powers a collimated (optical) outflow. 
Here we show the first maps of the molecular outflow associated
with RNO\,129.

\subsection{RNO\,91}
RNO\,91 and its source, IRAS\,16316-1540,
 are located in the L\,43 cloud, part of  the
 $\rho$~Oph molecular cloud complex at a distance of  
about 160 pc.  Optical, CO, CS, and NH$_3$ observations
of the core suggest that 
the molecular outflow has blown through the dense gas, creating
a cavity south of the outflow source (Mathieu et al.~1988). 
Based on single-dish and interferometer maps at higher angular resolution,
the molecular outflow has been 
successfully modeled as a slowly expanding shell
driven by a wide-angle wind (Bence et al.~1998; Lee et al.~2000, 2001;
Lee \& Ho 2005).
Given the large width of the outflow
near its base, about $10^4$ AU, our observations, 
which are limited to regions within about $6.5 \times 10^3$ AU of the source,
do not show the full impact of the RNO\,91 outflow on the surrounding 
envelope. Fortunately, the mosaic of interferometer observations 
by Lee \& Ho (2005), which cover  the surrounding environment 
within at least $10^4$~AU of IRAS\,16316-1540,
show these outflow-envelope interactions.

\subsection{T Tau}
The T Tau binary system comprises the optically visible star T Tau N and an infrared
companion, T Tau S, separated by only 0.7\arcsec . Both binary elements
are known outflow sources. The outflow mostly likely driven by T Tau S has 
an approximately north-west axis (e.g., Solf \& B\"ohm 1999;
Reipurth et al.~1997, and references therein), while the other seen
 nearly pole-on, with
a northeast-southwest or east-west axis, is most likely driven by T Tau N 
(Solf \& B\"ohm 1999; Momose et al.~1996,
and references therein). It is generally accepted that both T Tau N and T Tau S
are of similar age; T Tau N is optically visible because its light can escape
through the nearly pole-on outflow cavity walls (e.g., Momose et al.~1996),
while T Tau S is obscured by an envelope or the outer parts of T Tau N's disk,
and is highly reddened (e.g., Hogerheijde et al.~1997b). However, 
the bolometric temperatures
of the two stars are extremely different: 501 and 3452~K
for T Tau S and T Tau N, respectively (Chen et al.~1995).
Since we do not have separate bolometric 
temperature measurements for each element of other close ($< 5\arcsec$)
binary systems in our sample (i.e., IRAS\,3282, HH\,300, L\,1228, RNO\,129), and
since T Tau S and T Tau N should be coeval, we adopt a
 single  $T_{bol}$ for both stars using flux measurements 
 at different wavelengths that do not resolve the binary companions.

The molecular gas morphology and kinematics in the region surrounding
T Tau are quite complex, and different observational
studies offer different interpretations.
 Weintraub et al.~(1989b) conducted the first interferometer observations
 (with a beam of $\sim 6\arcsec$)
 of the \co(1-0) and \thco(1-0) surrounding T Tau  and interpreted their 
 observations to show a circumstellar Keplerian disk. The 
 combined
 interferometer and single-dish \thco(1-0)  observations  of 
 Momose et al.~(1996) show a   
 different picture, and were interpreted to show biconical
 outflowing shells. The interpretation by Schuster et al.~(1997)
 based on their single-dish \co(3-2) and \ceo(2-1)
 observations (with beams of 12 to 14\arcsec) is similar to that
 of Momose et al.~(1996), but they claim more than one
 outflow is needed to explain the observed gas morphology
 and kinematics. Observations of the extended 
 molecular hydrogen in the $v = 1 - 0$ S(1) line show the emission
 appears to arise from a precessing jet in an approximate north-south axis
 powered by T Tau S (van Langevelde et al.~1994b). Interferometer
 observations of the \hco(1-0) towards T Tau show possible evidence
 of a infalling circumstellar envelope (van Langevelde et al.~1994a).

The OVRO \co(1-0) maps presented here are much more sensitive, have a 
better angular resolution (by a factor of two), and better uv-coverage
(by a factor more than 4)  than those presented by 
Weintraub et al.~(1989b).  Our data does not show the Keplerian
circumstellar disk suggested by Weintraub et al.~(1989b). 
Instead, at velocities far away from the cloud velocity
(i.e., $v_{lsr} \la 6.5$~\kms \/ and $v_{lsr} \ga 9.5$~\kms) the \co \/
emission shows outflow lobes
with an east-west or northeast-southwest axis,  consistent 
with the Momose et al.~(1996) picture ---where most of the outflowing
molecular emission comes from an outflow powered by T Tau N.
Close to the cloud velocity the \co \/ emission (not shown here)
 is complex
and extended, and it is clear that we are missing a substantial amount
of flux. Our \thco , \ceo \/ and \hco \/ maps are consistent with the lower
angular resolution maps presented 
by Momose et al.~(1996), Schuster et al.~(1997), 
van Langevelde et al.~(1994a), respectively.

\subsection{GK/GI Tau}
GK Tau and GI Tau are two classical T Tauri stars located in the B18 cloud
in Taurus, that are only separated by 13\arcsec \/ ($\sim 1820$ AU). 
In addition, GK Tau has a faint companion only 2.5\arcsec \/ (350 AU) away
(Reipurth \& Zinnecker 1993). 
Estimates of the age and bolometric temperature 
for GI Tau and GK Tau indicate that these 
stars are coeval (Hartigan et al.~1994; Chen et al.~1995).  
Here we assume a common bolometric temperature
of 2700 K.
Aspin \& Reipurth (2000) detected several HH knots near the triple system,
and were able to determine that GK Tau (or its faint companion) is the driving
source of at least some of the HH knots.
Their observations also reveal that the entire triple 
system is surrounded by 
a common envelope of reflection nebulosity. 
Our CO maps of the region are the first to show the molecular outflow associated
with the GK/GI system.

\clearpage


 \figcaption{Gallery of $^{12}$CO(1-0) outflows.
Blue (red) contours show the 
blueshifted (redshifted) emission from the outflow lobes
of the sources in our sample,
integrated over the velocity ranges given in Table~\ref{velflowtab}.
The upper, middle and lower panels show the class 0, class I, and 
class II objects, respectively. 
A cross marks the  position of the protostar  given by
the millimeter continuum emission peak. In each map,  
the synthesized
beam is in the lower left corner.
A scale  equivalent to 10000~AU at the assumed
distance of each source is also given.
The arrowed dashed line in each panel
represents the presumed outflow axis.  The values of the first
contour and the subsequent contour steps for the blue (red) lobe 
of each outflow are given inside brackets in 
units of Jy~\kms , at the lower left (right) corner of each panel.  
\label{twcoouts}}


 \figcaption{Gallery of $^{13}$CO(1-0) outflows.
Solid blue (red) contours show the 
blueshifted (redshifted) integrated intensity. The dashed contours
show the \co \/ outflows (shown in Figure~\ref{twcoouts}). 
The velocity range of integration for each \thco \/ outflow
lobe is given in Table~\ref{velflowtab}. 
The cross symbol indicates the position of the protostar, given by
 the millimeter continuum emission peak. 
The synthesized beam of the \thco \/ (\co ) 
 map is shown on  the lower right (left) corner of each panel.
The scale in each panel shows the equivalent of 5000~AU at the assumed
source distance. The values of the first
contour and the subsequent contour steps for each outflow lobe
are given inside brackets in units of Jy~\kms . 
In the case of RNO\,43, the ``blueshifted'' lobe is shown in grey,
as the gas velocities are very close the ambient envelope
velocity. The RNO\,43 outflow axis is very close to the
plane of the sky, thus this \thco \/
structure most likely traces the 
walls of the outflow cavity of the
blueshifted outflow lobe, albeit 
its  velocities 
(see Paper II).
\label{thcoouts}}


\figcaption{Gallery of HCO$^+$(1-0) outflows. Solid blue (red)
contours represent integrated intensity of blueshifted (redshifted)
\hco \/ emission. Dashed contours represent \co \/ molecular outflow.
The synthesized beam of the \hco \/ (\co ) map is shown on  the 
lower right (left)  corner of each panel. The values of the first
contour and the subsequent contour steps for the blue (red) lobe 
of each outflow are given inside brackets in 
units of Jy~\kms , at the lower left (right) corner of each panel.  
\label{hcoout}}


\figcaption{\ceo \/ integrated intensity (solid green) contour maps superimposed
on \co \/ outflows (dashed contours). The velocity range of integration for each 
source is given in Table~\ref{envtab}. The synthesized
beam of the \ceo \/ (\co ) map is shown on  the lower right (left) corner of each panel.
The scale in each panel shows the equivalent of 5000~AU at the assumed
source distance. The values of the first
contour and the subsequent contour steps are given in brackets 
at the bottom of each panel. 
\label{ceocores}}


 \figcaption[f5.eps]{Outflow opening angle as a function of source age.  
Source age was obtained using the $T_{bol}$-age relation 
from Ladd et al.~(1998).
Filled circles show sources in our sample for which we were able
to measure an outflow opening angle
(see Tables~\ref{sourcestab}
and \ref{outtabi}), and
the filled triangles 
are sources from the literature (see Table~\ref{openangtab}). 
The dashed line 
represents the fit to all the data shown in the figure.
The errors in the opening angle come from the 
uncertainty of the angle measurement, while the 
errors in the age come mainly from the  $T_{bol}$-age relation.
\label{openang}}


 \figcaption{First moment maps of the \ceo (1-0) line from our class 0
(top row) and class I (bottom row) sources. 
The first moment between two velocities,
$v_1$ and $v_2$, is defined as $\int_{v_1}^{v_2} T_B v dv /
\int_{v_1}^{v_2} T_B  dv$, where $T_B$ is the brightness 
temperature. The color bar to the right of each panel shows
the color that represents the velocity in each map. 
Contours represent integrated intensity levels similar to those
of Figure~\ref{ceocores}.
The arrow shows the \co \/ outflow axis, from Figure~\ref{twcoouts}.
\label{envgrad}}


\figcaption{Envelope mass as a function of protostar age.
The envelope mass was obtained from the \ceo \/ emission 
(see Table~\ref{envtab}). 
The protostar age was obtained using the $T_{bol}$-age relation 
from Ladd et al.~(1998). The dashed line 
represents the fit to all the data shown in the figure.
The error bars in the envelope mass
reflect the 30\% calibration uncertainty in the data, while
the error in the age come mainly from the  $T_{bol}$-age relation.
\label{envmasfig}}


\figcaption{Schematic picture of outflow-envelope interaction evolution. 
Dark grey regions denote high-density (envelope) gas, 
mostly traced by \ceo \/ in our observations. Light
grey regions show the molecular outflow traced by the \co \/ 
and \thco . Arrows indicate the gas motion. 
See \S~\ref{outenv} for details.
\label{modelfig}}


\setcounter{table}{0}

\clearpage

\begin{deluxetable}{lccccccc}
\tablecolumns{8}
\tabletypesize{\small}
\tablecaption{Observed Sources
\label{sourcestab}}
\tablehead{
\colhead{Source Name} & \colhead{Alternate Name} &
\multicolumn{2}{c}{Observation Coordinates}
 & \colhead{d} &
\colhead{Class} & \colhead{$T_{bol}$\tablenotemark{a}} & \colhead{SED References\tablenotemark{b}}\\
  &  & \colhead{R.A. [2000]} & \colhead{Dec. [2000]} & \colhead{[pc]} &
\colhead{[K]} & 
}
\startdata
HH\,114mms & ... & 5$^h$18$^m$15$^s$.4 & 7\arcdeg 12\arcmin 00\arcsec & 460 & 0 & 40 & 1\\
IRAS 05295+1247 & RNO\,43 & 5 32 19.4 &  12 49 42 & 460 & 0 & 40 & 2,3\\
IRAS 03282+3035 & IRAS\,3282 & 3 31 21.0 & 30 45 31 & 350 & 0 & 50 & 2,4,5\\
IRAS 04239+2436 & HH\,300 & 4 26 56.3 & 24 43 36 & 140 & I & 225 & 2,5,6,7\\
IRAS 20582+7724 & L\,1228 & 20 57 13.0 & 77 35 47 & 200 & I & 345  & 2,7,8\\
IRAS 21004+7811 & RNO\,129 & 20 59 13.8 & 78 23 05 & 200 & I & 650  & 2,7\\
IRAS 16316-1540 & RNO\,91 & 16 34 29.3 & -15 47 01 & 160 & II & 715\tablenotemark{c} & 2,7,9\\
T Tau & ... & 4 21 59.4 & 19 32 07 & 140 & II & 1235 & 2,7,9,10,11,12,13\\
GK/GI Tau\tablenotemark{d} & ... &  4 33 34.4 & 24 21 07 & 140 & II & 2700\tablenotemark{c} & 7,9\\
\enddata  

\tablenotetext{a}{Bolometric temperature calculated from spectral energy distribution 
following Chen et al.~(1995) and combining our new observations with published data. 
Note that for T Tau, the flux measurements
used do not resolve the individual components (see A.8).}
\tablenotetext{b}{References used to obtain spectral energy distribution for each source:
 (1) Chini et al.~(1997);
 (2) IRAS Catalog of Point Sources; (3) Zinnecker et al.~(1992); (4) Dent et al.~1998; (5)  Motte \& Andr\'e (2001);
(6) Myers et al.~(1987); (7) 2MASS All-Sky Catalog of Point Sources; 
(8) Arce \& Sargent~(2004); (9) Herbig \& Bell (1988); (10) Weintraub et al.~(1987);
(11) Weintraub et al.~(1989a); (12) Beckwith et al.~(1990);  (13) Beckwith \& Sargent (1991) .}
\tablenotetext{c}{Bolometric temperature used here was taken directly from Chen et al.~(1995).} 
\tablenotetext{d}{GK Tau and GI Tau are less than 15\arcsec \/ 
apart and both are inside our field of view.}

\end{deluxetable}

\begin{deluxetable}{lcccccc}
\tablecolumns{7}
\tabletypesize{\small}
\tablecaption{Characteristics of Line and Continuum Maps
\label{linestab}}
\tablehead{
\colhead{Source} & 
\colhead{Line} &
\colhead{Beam Size} &
\colhead{Map RMS\tablenotemark{a}} &
\colhead{Line/Continuum} &
\colhead{Beam Size} &
\colhead{Map RMS\tablenotemark{a}}\\
 &  &
\colhead{[arcsec]} &
\colhead{[Jy~beam$^{-1}$]} &
&
\colhead{[arcsec]} &
\colhead{[Jy~beam$^{-1}$]}
}
\startdata
HH\,114mms & & & & & \\ 
 &  $^{12}$CO(1-0)  & $4.3 \times 3.9$ & 0.08 & HCO$^+$(1-0) & $6.4 \times 5.4$ & 0.11\\
 &  $^{13}$CO(1-0)  & $5.1 \times 5.0$ & 0.09 & 2.7~mm cont. & $6.4 \times 5.9$ & 0.003\\
 &  C$^{18}$O(1-0)  & $5.3 \times 4.9$ & 0.07 & 3.4~mm cont. & $6.3 \times 6.1$ & 0.0007\\
RNO\,43 &   &   &   &   &  \\ 
 &  $^{12}$CO(1-0)  & $4.4 \times 4.0$ & 0.10 & HCO$^+$(1-0) & $5.4 \times 5.2$ & 0.13\\
 &  $^{13}$CO(1-0)  & $5.1 \times 4.1$ & 0.10 & 2.7~mm cont. & $5.1 \times 5.0$ & 0.0007\\
 &  C$^{18}$O(1-0)  & $4.7 \times 4.1$ & 0.08 & 3.4~mm cont. & $5.6 \times 4.6$ & 0.0004\\
 IRAS\,3282 &   &   &   &   &  \\
 &  $^{12}$CO(1-0)  & $3.5 \times 2.9$ & 0.06 & HCO$^+$(1-0) & $6.4 \times 5.9$ & 0.13\\
 &  $^{13}$CO(1-0)  & $4.9 \times 4.1$ & 0.08 & 2.7~mm cont. & $4.3 \times 4.0$ & 0.001\\
 &  C$^{18}$O(1-0)  & $3.8 \times 3.4$ & 0.06 & 3.4~mm cont. & $5.0 \times 4.0$ & 0.0006\\
HH\,300 &   &   &   &   &  \\
 &  $^{12}$CO(1-0)  & $4.4 \times 3.2$ & 0.07 & HCO$^+$(1-0) & $6.9 \times 6.6$ & 0.11\\
 &  $^{13}$CO(1-0)  & $4.6 \times 3.3$ & 0.10 & 2.7~mm cont. & $4.8 \times 3.7$ & 0.0008\\
 &  C$^{18}$O(1-0)  & $4.7 \times 3.5$ & 0.06 & 3.4~mm cont. & $4.8 \times 3.8$ & 0.0004\\
L\,1228 &   &   &   &   &  \\
 &  $^{12}$CO(1-0)  & $4.3 \times 3.8$ & 0.08 &  HCO$^+$(1-0) & $7.1 \times 6.6$ & 0.11\\
 &  $^{13}$CO(1-0)  & $4.8 \times 4.0$ & 0.07 &  2.7~mm cont. & $6.0 \times 5.7$ & 0.0006\\
 &  C$^{18}$O(1-0)  & $4.7 \times 3.9$ & 0.08 &  3.4~mm cont. & $6.3 \times 5.5$ & 0.0004\\
RNO\,129 &   &   &   &   &  \\
 &  $^{12}$CO(1-0)  & $5.7 \times 4.4$ & 0.11 &  HCO$^+$(1-0) & $8.5 \times 5.3$ & 0.10\\
 &  $^{13}$CO(1-0)  & $6.4 \times 5.0$ & 0.10 &  2.7~mm cont. & $7.9 \times 6.9$ & 0.002\\
 &  C$^{18}$O(1-0)  & $6.8 \times 5.7$ & 0.08 &  3.4~mm cont. & $9.1 \times 6.0$ & 0.0007\\
RNO\,91 &   &   &   &   &  \\
 &  $^{12}$CO(1-0)  & $6.2 \times 4.5$ & 0.16 &  HCO$^+$(1-0) & $11.7 \times 6.3$ & 0.25\\
 &  $^{13}$CO(1-0)  & $6.1 \times 4.9$ & 0.15 &  2.7~mm cont. & $6.6 \times 4.7$ & 0.001\\
 &  C$^{18}$O(1-0)  & $6.6 \times 5.0$ & 0.12 &  3.4~mm cont. & $13.3 \times 7.8$ & 0.0009\\
T Tau &   &   &   &   &  \\
 &  $^{12}$CO(1-0)  & $3.4 \times 3.0$ & 0.08 &  HCO$^+$(1-0) & $5.2 \times 4.5$ & 0.12\\
 &  $^{13}$CO(1-0)  & $3.7 \times 3.6$ & 0.09 &  2.7~mm cont. & $4.6 \times 4.3$ & 0.002\\
 &  C$^{18}$O(1-0)  & $4.3 \times 3.8$ & 0.06 &  3.4~mm cont. & $4.9 \times 4.1$ & 0.0005\\
GK Tau &   &   &   &   &  \\
 &  $^{12}$CO(1-0)  & $5.9 \times 4.4$ & 0.10 &  HCO$^+$(1-0) & $5.3 \times 4.1$ &  0.16\\
 &  $^{13}$CO(1-0)  & $7.0 \times 6.1$ & 0.12 &  2.7~mm cont. & $7.1 \times 5.3$ & 0.0008\\
 &  C$^{18}$O(1-0)  & $6.9 \times 5.7$ & 0.09 &  3.4~mm cont. & $6.4 \times 6.1$ & 0.0004\\ 

\enddata  
\tablenotetext{a}{In line maps, the rms per velocity channel is shown.}
\end{deluxetable}

\begin{deluxetable}{lllccccc}
\tablecolumns{8}
\tabletypesize{\footnotesize}
\tablecaption{Continuum Results 
\label{contab}}
\tablehead{
\colhead{Source} & \colhead{R.A.} & \colhead{Dec.} &
\colhead{$F_{2.7}$\tablenotemark{a}} & \colhead{$F_{3.4}$\tablenotemark{a}} 
& \colhead{$\beta$\tablenotemark{b}} &
\colhead{$T_d$\tablenotemark{c}} & \colhead{$M_{cont}$\tablenotemark{d}}\\
  &  &  &  
   \colhead{[mJy]} &  \colhead{[mJy]} &  &  
  \colhead{[K]} & \colhead{[$M_{\sun}$]} 
}
\startdata
HH\,114mms & 5$^h$18$^m$15$^s$.2 & 7\arcdeg 12\arcmin 02\arcsec & $110 \pm 20$ & $57 \pm 11$ & $0.4 \pm 0.2$ & 30 & 0.24\\
RNO\,43 & 5 32 19.4 &  12 49 41 & $26 \pm 6$ & $11 \pm 2$ & $1.0 \pm 0.1$ & 35  & 0.19\\
IRAS\,3282 & 3 31 20.9 & 30 45 30  &  $61 \pm 12$ & $30 \pm 6$ & $0.76 \pm 0.2$ &  28 & 0.17\\
HH\,300 & 4 26 56.3 & 24 43 35 & $14 \pm 2$ & $8 \pm 2$ & $1.5 \pm 0.4$ &40 & 0.02\\
L\,1228 & 20 57 12.9 & 77 35 44 & $45 \pm 9$ & $21 \pm 4$ & $1.2 \pm 0.4$ & 33  &  0.10\\
RNO\,129 & 20 59 13.6 & 78 23 03 & $62 \pm 13$ & $28 \pm 6$ & 1.0 & 40 & 0.07\\
RNO\,91 & 16 34 29.3 & -15 47 01 & $8 \pm 2$ & $3 \pm 0.5$ & 1.0 & 50 & 0.005\\
T Tau & 4 21 59.5 & 19 32 06 & $60 \pm 12$ & $33 \pm 6$ & $0.25 \pm 0.2$ & 45 & 0.005\\
GK/GI Tau\tablenotemark{e} & 4 33 34.53/34.13 & 24 21 06/18 & $<2.0$\tablenotemark{f}/$4.0 \pm  0.8$ & $1.0 \pm 0.2$/$3.0 \pm 0.6$ & 1.0 & 40 & 0.001/0.003\\ 
\enddata  

\tablenotetext{a}{Measured continuum fluxes at 2.7~mm ($F_{2.7}$), and at 3.4~mm ($F_{3.4}$).}
\tablenotetext{b}{We show the 1-$\sigma$ error of our estimate of 
$\beta$, the power-law index of the dust emissivity. We assigned $\beta = 1$ 
to sources for which we do not have enough data to estimate its value 
(see text for details).}
\tablenotetext{c}{Uncertainties in $T_d$ are about 20\%. }
\tablenotetext{d}{Uncertainties in $M_{cont}$ are about 30\%. }
\tablenotetext{e}{The dust continuum emission associated with GI Tau
(a companion star to GK Tau) is also detected in the GK Tau field}
\tablenotetext{f}{Quoted value is 3-$\sigma$ upper limit.}

\end{deluxetable}

\begin{deluxetable}{lllllll}
\tablecolumns{8}
\tabletypesize{\footnotesize}
\tablecaption{Velocity Ranges of Molecular Outflow Lobes  
\label{velflowtab}}
\tablehead{
\colhead{Source} & 
\multicolumn{2}{c}{\co \/ Outflow} & 
\multicolumn{2}{c}{\thco \/ Outflow} & \multicolumn{2}{c}{\hco \/ Outflow} \\
& \colhead{Blue Lobe} & \colhead{Red Lobe}
& \colhead{Blue Lobe} & \colhead{Red Lobe}
& \colhead{Blue Lobe} & \colhead{Red Lobe}\\
& \colhead{\kms} & \colhead{\kms} & \colhead{\kms} & \colhead{\kms} & \colhead{\kms} & \colhead{\kms}
}
\startdata
HH\,114mms  & [-10.05, -3.87]  & [0.36, 8.16] & [-1.26, -1.08] & [0.10, 0.62] & ... & [0.19, 1.23]\\
RNO\,43        &  [6.02, 8.95] & [11.22, 16.10] & [10.11, 10.45]\tablenotemark{a} & [10.63, 11.14] & ... & [10.76, 11.17]\\
IRAS\,3282    &  [-2.75, 5.05] &  [8.63, 16.10] & [5.98, 6.49] & [7.18, 7.52] & [4.28, 6.05] & [7.11, 9.52] \\
HH\,300         &  [3.45, 6.05] & [7.03, 10.28] & [5.37, 6.02] & [7.04, 7.90] & [5.86, 6.07] & ...\\
L\,1228          & [-17.50, -9.05] &  [-6.13, 2.00] & [-9.63, -8.60] & [-6.90, -6.40] & [-10.90, -9.02] & [-6.92, -5.02]\\
RNO\,129    & [-17.75, -10.93] & [-6.38, -2.14] & [-9.87, -8.17]  & [-6.63, -5.79] & [-8.84, -8.21] & [-6.63,-5.90]\\
RNO\,91      &  [-6.65, -0.80]  & [1.15, 2.13] & [-0.70, -0.02] & [1.01, 1.34] & [-0.34, -0.10] & ... \\ 
T Tau          &  [1.17, 6.70] & [9.30, 13.53] & [5.44, 7.14] & [8.52, 10.05] & [5.02, 7.19] & [8.81, 9.81]  \\
GK/GI Tau  & [4.95, 6.25] & [7.55, 9.50] & [5.87, 6.38] & [7.42, 7.92] & ... & ...\\
\enddata
\tablenotetext{a}{Although the entire range of velocities is not blueshifted with respect to the RNO\,43
ambient envelope velocity, 10.2~\kms , we are convinced the \thco \/ structure at these velocities 
traces the outflow cavity walls of the RNO\,43 blueshifted outflow (see Paper II).}

\end{deluxetable}

\begin{deluxetable}{lccccc}
\tablecolumns{6}
\tabletypesize{\footnotesize}
\tablecaption{Molecular Outflow Parameters I
\label{outtabi}}
\tablehead{
\colhead{Source} & 
\colhead{Outflow Axis PA} & \colhead{Opening Angle\tablenotemark{a}} &
 \colhead{$M_{out}$} & \colhead{$P_{out}$\tablenotemark{b}} & 
\colhead{$E_{out}$\tablenotemark{b}} \\
 &  \colhead{(deg. E of N)}  &  \colhead{[deg.]} 
 & \colhead{($10^{-3}$ M$_{\sun}$)} 
 & \colhead{($10^{-3}$ M$_{\sun}$~\kms )} 
 & \colhead{($10^{40}$ erg)} 

}
\startdata
HH\,114mms & 60 & 25 & 30 & 91 & 484\\
RNO\,43 & 60 & 45 & 150 & 119 & 142\\
IRAS\,3282 & 125 & 50 & 15 & 44 & 195\\
HH\,300 & 59 &  80 & 8 & 6 & 2\\
L\,1228 & 79 & 95 & 13 & 26 & 92\\
RNO\,129 & 95 & 125 & 24 & 36 & 94\\
RNO\,91 & 155 & 160 & 11 & 11 & 20\\ 
T Tau & 75 & 100 & 21 & 25 & 37\\
GK/GI Tau & ...\tablenotemark{c} &  ... & 6 & 4 & 4\\
\enddata

\tablenotetext{a}{Outflow opening angle was estimated the following way:
in HH\,114mms we used the lowest velocity blueshifted \co ; in RNO\,43 
and L\,1228 we 
used the blushifted \thco \/ outflow lobe shown in Figure~\ref{thcoouts};
in IRAS\,3282, HH\,300, RNO\,129, and TTau
we used both lobes of the \co \/ outflow shown in Figure~\ref{twcoouts};
in RNO\,91 we used the opening angle estimate from Lee \& Ho (2005);
in GK/GI Tau we unable to determine the opening angle
from our maps.}
\tablenotetext{b}{Estimated value not corrected for outflow inclination with
respect to the plane of the sky}
\tablenotetext{c}{Unknown outflow axis due to uncertainty as to which source 
powers the observed outflow.}

\end{deluxetable}

\begin{deluxetable}{lccc}
\tablecolumns{4}
\tabletypesize{\footnotesize}
\tablecaption{Molecular Outflow Parameters II
\label{outtabii}}
\tablehead{
\colhead{Source} & 
\colhead{$\dot{M}_{out}$\tablenotemark{a}} & \colhead{$F_{co}$\tablenotemark{a}} &
\colhead{$\dot{M}_{dense}$\tablenotemark{a}}\\
 &  
 \colhead{($10^{-7}$ M$_{\sun}$~yr$^{-1}$)} 
& \colhead{($10^{-6}$ M$_{\sun}$~\kms \/ yr$^{-1}$)}
&  \colhead{($10^{-7}$ M$_{\sun}$~yr$^{-1}$)}
}
\startdata
HH\,114mms & 41 & 16 & 11\\
RNO\,43 & 73 & 9 & 51\\
IRAS\,3282 & 31 & 11 & 4\\
HH\,300 & 18 & 1 & 16\\
L\,1228 & 29 & 15 & 13\\
RNO\,129 & 56  & 14 & 37\\
RNO\,91 & 16 & 3 & 11\\ 
T Tau & 47  & 8 & 36\\
GK/GI Tau & 2 & 0.2 & 2\\
\enddata
\tablenotetext{a}{Estimated value not corrected for outflow inclination with
respect to the plane of the sky}

\end{deluxetable}

\begin{deluxetable}{lcc}
\tablecolumns{3}
\tabletypesize{\small}
\tablecaption{\hco \/ Abundance in Outflows 
\label{hcotab}}
\tablehead{
\colhead{Source} & 
\multicolumn{2}{c}{[HCO$^+$]/[H$_2$]} \\
& \colhead{Blue Lobe} & \colhead{Red Lobe} \\
& \colhead{($10^{-9}$)} & \colhead{($10^{-9}$)}
}
\startdata

HH\,114mms  & ...     & $> 20$ \\
RNO\,43    & ...     & $> 2$  \\
IRAS\,3282  & $> 2$  & $> 60$  \\
HH\,300    & 5       &    1   \\  
L\,1228     & $> 7$   & $> 4$  \\
RNO\,129   &  0.5    & 0.6    \\
RNO\,91    &  1      & ...    \\ 
T Tau     &  4      &  3     \\
GK/GI Tau & ...     & ...    \\
\enddata
\end{deluxetable}

\begin{deluxetable}{llccccc}
\tablecolumns{7}
\tabletypesize{\small}
\tablecaption{\ceo (1-0) Envelope Parameters 
\label{envtab}}
\tablehead{
\colhead{Source} & \colhead{Velocity Range\tablenotemark{a}}
& \colhead{Size\tablenotemark{b}} & \colhead{P.A.\tablenotemark{c}} & 
\colhead{$M_{env}$} & \colhead{$v_{env}$} & \colhead{Reference\tablenotemark{d}}\\
 & \colhead{(\kms )} & \colhead{(AU)}
& \colhead{(deg. E of N)} & \colhead{($M_{\sun}$)}
 & \colhead{(\kms )}  &
}
\startdata

HH\,114mms &  $[-1.09, 0.28]$ & $2700 \times 1400$ & 69 & 0.13 
            & -0.46 & 1\\
RNO\,43 & $[9.26, 11.48]$ & $6200 \times 5000$ & 75 & 0.87 
            & 10.20 & 1\\
IRAS\,3282 &  $[6.32, 7.68]$ & $1900 \times 1600$ & 113 & 0.08 
            & 6.96 & 1\\
HH\,300 & $[5.84, 7.21]$ & $3700 \times 1300$ & 161 & 0.03 
            & 6.50 & 2\\
L\,1228 & $[-8.77, -7.41]$ &  $2400 \times 1500$ & 161 & 0.05 
           & -8.00 & 1\\
RNO\,129 & $[-8.35, -6.29]$ & $2000 \times 1500$ & 149 & 0.04 
           & -7.60 & 1\\
RNO\,91 & $[-0.35, 0.33]$  & ... & ... & 0.01 
           & 0.57 & 1\\
T Tau & $[7.15, 8.34]$ & ... & ... & 0.02 
          & 7.90 & 3\\
GK/GI Tau & ... & ... & ... & $<0.001$\tablenotemark{e} 
          & 6.90 & 4\\
\enddata

\tablenotetext{a}{\phantom{}Lower and upper limits of velocity range for 
which \ceo \/ emission was detected above the 3-$\sigma$ level.}
\tablenotetext{b}{FWHM size of \ceo \/ envelope from a gaussian fit to the 
integrated intensity map, deconvolved by the beam.}
\tablenotetext{c}{Deconvolved position angle of the gaussian fitted to the \ceo \/
integrated intensity emission.}
\tablenotetext{d}{References for estimate of $v_{env}$: (1) our observations; 
(2) Fuller \& Ladd (2002); (3) average of 
values used by Momose et al.~(1996) and van Langevelde et al.~(1994a); (4)
peak velocity at position of GK Tau from Mizuno et al.~(1995).}
\tablenotetext{e}{Quoted value is 3-$\sigma$ upper limit.}
\end{deluxetable}

\begin{deluxetable}{lcccc}
\tablecolumns{5}
\tabletypesize{\small}
\tablecaption{Outflow Opening Angle from Literature  
\label{openangtab}}
\tablehead{
\colhead{Source} & \colhead{T$_{bol}$}  & \colhead{Reference\tablenotemark{a}}
& \colhead{Opening Angle\tablenotemark{b}} & \colhead{Reference\tablenotemark{c}}\\ 
 & \colhead{(K)} &  & \colhead{(deg)} & 
}
\startdata

IRAM\,04191 &   18   & 1 &  45  &  1\\
HH\,211  &           30   & 1 & 15   &  2\\        
L\,1448   &            56  &  1 & 40  &   3\\  
L\,1527   &            59  &  1 & 70   &   4\\       
L\,1157   &            62  &  1 & 35   &    5\\      
HH\,114/115  &    65  &  2,3,4   & 30   & 1\\ 
RNO\,40    &         65  &  2,3,4 &   30  & 6\\ 
HH\,111   &        70  &  2,3,4  & 20   & 6\\
L\,1551-IRS5  &   97  & 1 & 100   & 7\\
B\,5-IRS1  &        185  & 4,5,6 & 90 & 8\\ 
\enddata
\tablenotetext{a}{T$_{bol}$ obtained from estimates shown in 
(1) Buckle \& Fuller (2002), or calculated
from SED using data from the literature: 
(2) Reipurth et al.~(1993); 
(3) Dent et al.~(1998); 
(4) IRAS Catalog of Point Sources; 
(5) 2MASS All-Sky Catalog of Point Sources; 
(6) Langer et al.~(1996).}
\tablenotetext{b}{Uncertainty in opening angle estimate is about
 5 to 10\arcdeg .}
\tablenotetext{c}{Reference for opening angle estimate: 
(1) Lee et al.~(2002); 
(2) Gueth \& Guilloteau (1999);
(3) Bachiller et al.~(1995); 
(4) Ohashi et al.~(1997); 
(5) Gueth et al.~(1996); 
(6) Lee et al.~(2000); 
(7) Ohashi et al.~(1996);
(8) Langer et al.~(1996).}

\end{deluxetable}

\end{document}